# Stretching Epitaxial La$_{0.6}$Sr$_{0.4}$CoO$_{3-\delta}$ for Fast Oxygen Reduction

Dongkyu Lee,[†] Ryan Jacobs,[‡] Youngseok Jee,[§] S. S. Ambrose Seo,[¶] Changhee Sohn,[†] Anton V. Ievlev,[†] Olga S. Ovchinnikova,[†] Kevin Huang,[§] Dane Morgan,[‡] and Ho Nyung Lee[†,*]

[†] *Oak Ridge National Laboratory, Oak Ridge, Tennessee 37831, USA*

[‡] *Department of Materials Science and Engineering, University of Wisconsin Madison, Madison, Wisconsin 53706, USA*

[§] *Department of Mechanical Engineering, University of South Carolina, Columbia, South Carolina 29208, USA*

[¶] *Department of Physics and Astronomy, University of Kentucky, Lexington, Kentucky 40506, USA*

*This manuscript has been authored by UT-Battelle, LLC under Contract No. DE-AC05-00OR22725 with the U.S. Department of Energy. The United States Government retains and the publisher, by accepting the article for publication, acknowledges that the United States Government retains a non-exclusive, paid-up, irrevocable, world-wide license to publish or reproduce the published form of this manuscript, or allow others to do so, for United States Government purposes. The Department of Energy will provide public access to these results of federally sponsored research in accordance with the DOE Public Access Plan (http://energy.gov/downloads/doe-public-access-plan).*

ABSTRACT

The slow kinetics of the oxygen reduction reaction (ORR) is one of the key challenges in developing high performance energy devices, such as solid oxide fuel cells. Straining a film by growing on a lattice-mismatched substrate has been a conventional approach to enhance the ORR activity. However, due to the limited choice of electrolyte substrates to alter the degree of strain, a systematic study in various materials has been a challenge. Here, we explore the strain modulation of the ORR kinetics by growing epitaxial $La_{0.6}Sr_{0.4}CoO_{3-\delta}$ (LSCO) films on yttria-stabilized zirconia substrates with the film thickness below and above the critical thickness for strain relaxation. Two orders of magnitude higher ORR kinetics is achieved in an ultra-thin film with ~0.8% tensile strain as compared to unstrained films. Time-of-flight secondary ion mass spectrometry depth profiling confirms that the Sr surface segregation is not responsible for the enhanced ORR in strained films. We attribute this enhancement of ORR kinetics to the increase in oxygen vacancy concentration in the tensile-strained LSCO film owing to the reduced activation barrier for oxygen surface exchange kinetics. Density functional theory calculations reveal an upshift of the oxygen 2$p$-band center relative to the Fermi level by tensile strain, indicating the origin of the enhanced ORR kinetics.

INTRODUCTION

Development of novel catalysts for the oxygen reduction reaction (ORR) and oxygen evolution reaction (OER) is highly desirable to fulfill the increasing demand for renewable energy materials and systems,[1] including oxygen permeation membranes,[2] air batteries,[3] and solid oxide fuel cells (SOFCs) and electrolyzers.[4] In such energy devices, transition metal perovskite oxides are widely used to catalyze the ORR and OER. Among various technical obstacles, slow ORR kinetics is known to be a critical rate-limiting step, reducing the efficiency of electrochemical devices. Recently, the use of epitaxial strain in a thin film induced by lattice mismatch with a substrate showed a possibility to control the electronic structure,[5] oxygen transport,[6] and oxygen defect formation,[7] ultimately enhancing the high temperature oxygen electrocatalysis in perovskite oxides.[8-10]

While substantial effort has been devoted to enhance the catalytic activity by strain, there is no systematic understanding of the underlying mechanism for the enhanced oxygen electrocatalytic activities by strain at elevated temperatures. Owing to concurrent changes of multiple material properties by strain, including the oxygen vacancy concentration, surface cation distribution, and the energy barrier for the surface exchange processes, it is difficult to deconvolute the main contributor from multiple factors induced by strain to the enhanced ORR activity. More importantly, there are only a limited number of available substrates to induce strain for high temperature electrochemical applications. For example, yttria-stabilized zirconia (YSZ) is the only single crystalline substrate available in a large size, satisfying both the growth (i.e., lattice mismatch) and electrochemical requirements (i.e., ionic conductor and electronic insulator). The limited choice of substrates imposes a constraint on the range of lattice strains explored in previous studies,[5, 8] which suggests the need to develop a new strategy to modulate the epitaxial strain for designing catalytically active materials.

An alternative approach to manipulate the strain is to control the film thickness instead of varying the lattice constant of the substrate by using different substrates (i.e., $SrTiO_3$ and $LaAlO_3$). While $La_{1-x}Sr_xCoO_{3-\delta}$ (x = 0, 0.2, and 0.4) thin films demonstrated enhanced oxygen surface exchange kinetics compared to bulk samples,[9, 11-12] films of varying thickness exhibited little or no strain variation. As a result, these films showed little dependence of their surface exchange kinetics and oxygen vacancy concentration as a function of their thickness. This lack of variation could be attributed to the fact that the critical thickness to maintain a certain strain state is smaller than what was used in the previous studies. Here, by growing ultrathin films with highly preserved epitaxial strain, we show that the change in the thickness of epitaxial $La_{0.6}Sr_{0.4}CoO_{3-\delta}$ (LSCO) films can also modulate the epitaxial strain, and thereby produces significantly enhanced oxygen electrocatalysis. While the thickness control is a simple means to tune strain, the enhancement in the ORR kinetics by controlling film thickness is found to be significant.

EXPERIMENTAL AND THEORETICAL METHODS

To systematically introduce strain in LSCO, epitaxial thin films with various thicknesses of 10, 25, 50, 80, and 100 nm were grown by pulsed laser epitaxy (PLE) on single crystal (001) yttria-stabilized zirconia (YSZ) substrates, which serve as both substrate and electrolyte. A Gd-doped ceria (GDC) buffer layer (4 nm in thickness) was introduced between YSZ and LSCO to prevent the formation of an unwanted $La_2Zr_2O_7$ layer at the interface.[13] Prior to LSCO and GDC deposition, platinum ink (Pt) counter electrodes were painted on one side of the YSZ and dried at 900 ºC in air for 1 hour. The YSZ substrate was affixed to the PLE substrate holder using a small amount of silver paint for thermal contact. PLE was performed using a KrF excimer laser at $\lambda$ = 248 nm, 10 Hz pulse rate and 50 mJ pulse energy under an oxygen partial pressure, $p(O_2)$ of 6.6 × $10^{-5}$ atm (50 mTorr) with GDC at 600 °C, followed by LSCO at 700 °C. After completing the LSCO deposition, the samples were cooled down to room temperature in the PLE chamber for ≈1 hour under a $p(O_2)$ of 6.6 × $10^{-5}$ atm (50 mTorr). All films showed an atomically flat surface with a root-mean-square roughness value below 0.5 nm revealed by atomic force microscopy (AFM) (see Figure S1 in the Supporting Information).

*In-situ* HRXRD was conducted on a four-circle diffractometer (Panalytical) equipped with a controlled temperature stage (DHS 900, Anton Paar) in a $p(O_2)$ of 1 atm (flux ≈5 sccm). Silver paste was used to adhere the thin film sample with the YSZ single crystal to the heating plate. Starting at 25 °C, a heating rate of ≈10 °C/min was chosen and after increasing the temperature, the temperature was always held constant for 30 minutes to reach thermal equilibrium before XRD data were collected. Sample realignment was conducted after each temperature step to optimize the intensity of the YSZ 002 peak for all out-of-plane scans. This procedure includes the correction of slight re-adjustments due to the thermal expansion of the sample (height offset and angles).

*In-situ* EIS measurements were conducted on asymmetrical cell structures containing the LSCO films with gold mesh current collectors fabricated by photolithography and sputtering.[14] EIS measurements were performed in the frequency range of 1 mHz to 1 MHz with AC amplitude of 10 mV and zero DC bias at temperatures between 350 °C and 650 °C, using the Frequency Response Analyzer module (FRA, Solartron 1260).

Depth profiling was carried out on a ToF.SIMS 5 instrument (ION-TOF GmbH, Germany). As primary ions, $Bi_3^+$ clusters were used with a beam voltage, current, and area of 30 keV, 30 nA, and 150 × 150 $\mu m^2$, respectively. Cs$^+$ ions were utilized in the sputter gun for depth profiling with a beam voltage, current, and area of 1 keV, 60 nA, and 300 × 300 $\mu m^2$, respectively.

*In-situ* ellipsometry technique was employed to obtain the optical constants of thin films in the same condition with EIS testing. By using M-2000 ellipsometer (J. A. Woollam Co.), two ellipsometric parameters, $\psi$ (intensity ratio between *p*- and *s*-polarization light) and $\Delta$ (phase difference between *p*- and *s*-polarization light) for the films and the bare substrate were obtained at 650 °C in 1 atm. We directly transformed ellipsometric parameters of the bare substrate to real and imaginary parts of optical conductivity. Then, we constructed a two-layer model composed of the substrate and film to determine the optical conductivity of the films. We fixed the thickness of the films as well as the optical conductivity of the substrate, and only varied the optical conductivity of the films during the fitting procedure. The Kramers-Kronig condition was forced during the procedure.

All calculations were performed using DFT within the Vienna Ab-Initio Simulation Package (VASP) with a plane wave basis set.[15] We used the generalized gradient approximation exchange and correlation functional with Hubbard *U* correction (GGA+*U*)[16] with the projector augmented wave (PAW) method[17] and pseudopotentials of Perdew and Wang (PW-91).[18] The Hubbard *U* correction was applied to Co atoms only, with $U_{eff}$ = 3.32 eV ($U$ = 3.32 eV and $J$ = 0 eV).[19] The valence electron configurations for La, Sr, Co and O elements used were: La: $5s^25p^66s^25d^1$, Sr: $4s^24p^65s^2$, Co: $4s^13d^8$, O: $2s^22p^4$. All calculations were performed with spin polarization enabled. The spin state of Co in all calculations was intermediate spin. While Co-containing perovskites have been shown to exist in a variety of spin states under different conditions, our usage of intermediate spin Co is justified for applications of oxygen electrocatalysis at elevated temperatures because the high temperature magnetic state of Co has been shown to be either intermediate spin or a spin mixture which is representative of an intermediate spin state.[12, 20-21] Reciprocal space integration in the Brillouin zone was conducted with the Monkhorst-Pack scheme[22] with 2×2×1 k-point sampling. $La_{0.6}Sr_{0.4}CoO_3$ cells were modeled as the high-temperature pseudocubic variant of the experimental rhombohedral phase (space group $R\bar{3}c$, number 167) with 100 atoms/cell following previous modeling work on perovskites for high temperature applications.[23-24] For the calculations involving strained cells, the strain was imposed on the *a*-axis and *b*-axis lattice parameters, and the *c*-axis was allowed to fully relax. The calculation of the oxygen *p* band center is performed using the methods detailed in previous works on DFT modeling of perovskites.[23, 25] As a function of strain, the oxygen *p* band centers of LSCO were calculated for five different Sr orderings, which were constructed by randomly choosing La sites to substitute for Sr while maintaining an A-site Sr concentration of 40% (see Figure S11 in the Supporting Information).

RESULTS AND DISCUSSION

X-ray diffraction (XRD) $\theta$-$2\theta$ scans at room temperature clearly revealed only 00*l* peaks from LSCO, GDC, and YSZ, indicating *c*-axis oriented films (Figure 1a). Similar to previous studies,[9, 11, 26] YSZ substrates with GDC buffer layers were found to introduce in-plane tensile strain (up to 1.6%) into LSCO thin films (see Figure S4 in the Supporting Information). Note that perovskite oxides often reveal a gradual strain relaxation as the film thickness is increased, which is most probably due to the formation of complex defect structures (such as dislocations) at the film/substrate interface to relieve the built-up elastic energy resulting from the strain. Such a phenomenon is also observed in this work, providing more than one strain state through use of multiple film thicknesses.

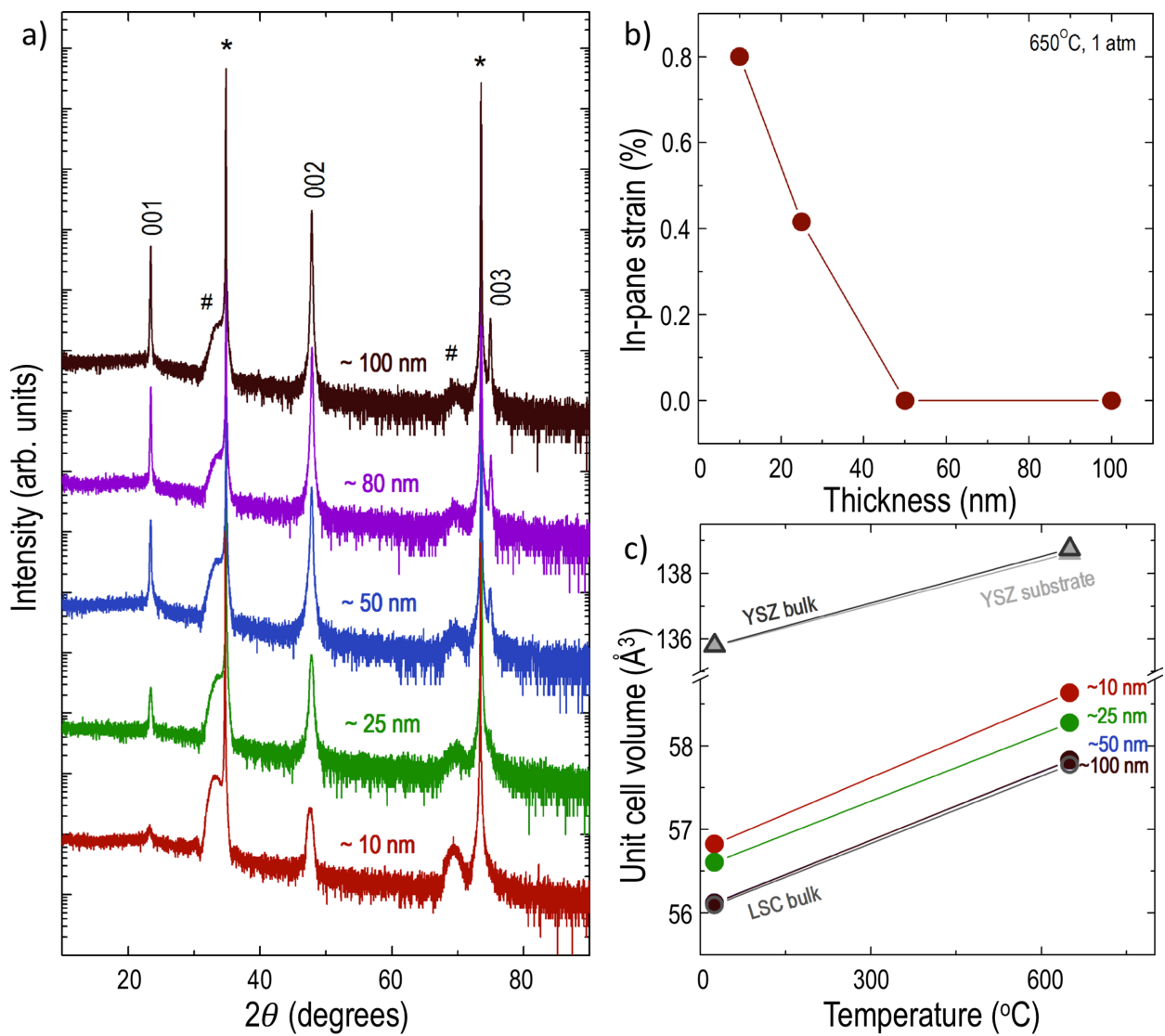

Figure 1. (a) XRD $\theta$-$2\theta$ patterns of LSCO thin films with thicknesses of 10, 25, 50, 80, and 100 nm measured at room temperature. Gd-doped ceria (GDC) and YSZ substrate peaks are indicated with # and *, respectively. (b) In-plane biaxial strain in the LSCO films as a function of film thickness obtained from high temperature XRD measurements at 650 °C in 1 atm, which is the condition we used for EIS measurements. (c) Unit cell volume change as a function of temperature. Reference bulk LSCO and YSZ unit cell volumes are also plotted for comparison.[27-28]

In order to check the strain state of our samples at the same condition for ORR testing, we also examined the evolution of the in-plane strain in LSCO films by performing XRD $\theta$-$2\theta$ scans at 650 °C in 1 atm (Figure 1b). Details on the strain calculations are provided in the Supporting

Information. It is worth noting that the in-plane tensile strain in LSCO thin films gradually decreased with increasing film thickness up to 50 nm, and no strain was built up in thicker films (≥ 50 nm). As represented in Figure 1c, the volumetric thermal expansion coefficients of our LSCO thin films (~15.5 × $10^{-6}$ °$C^{-1}$) and the YSZ substrate (~11.1 × $10^{-6}$ °$C^{-1}$) from XRD measurements were in good agreement with previously reported values (15.8 × $10^{-6}$ °$C^{-1}$ for bulk LSCO[29] and 11 × $10^{-6}$ °$C^{-1}$ for bulk YSZ[30]). Interestingly, the unit cell volume of LSCO thin films was found to increase with decreasing the film thickness, i.e., increasing in-plane tensile strain, which is consistent with the previous observation in $SrCoO_{3-\delta}$ thin films.[31] The origin of the increased unit cell volume is discussed later.

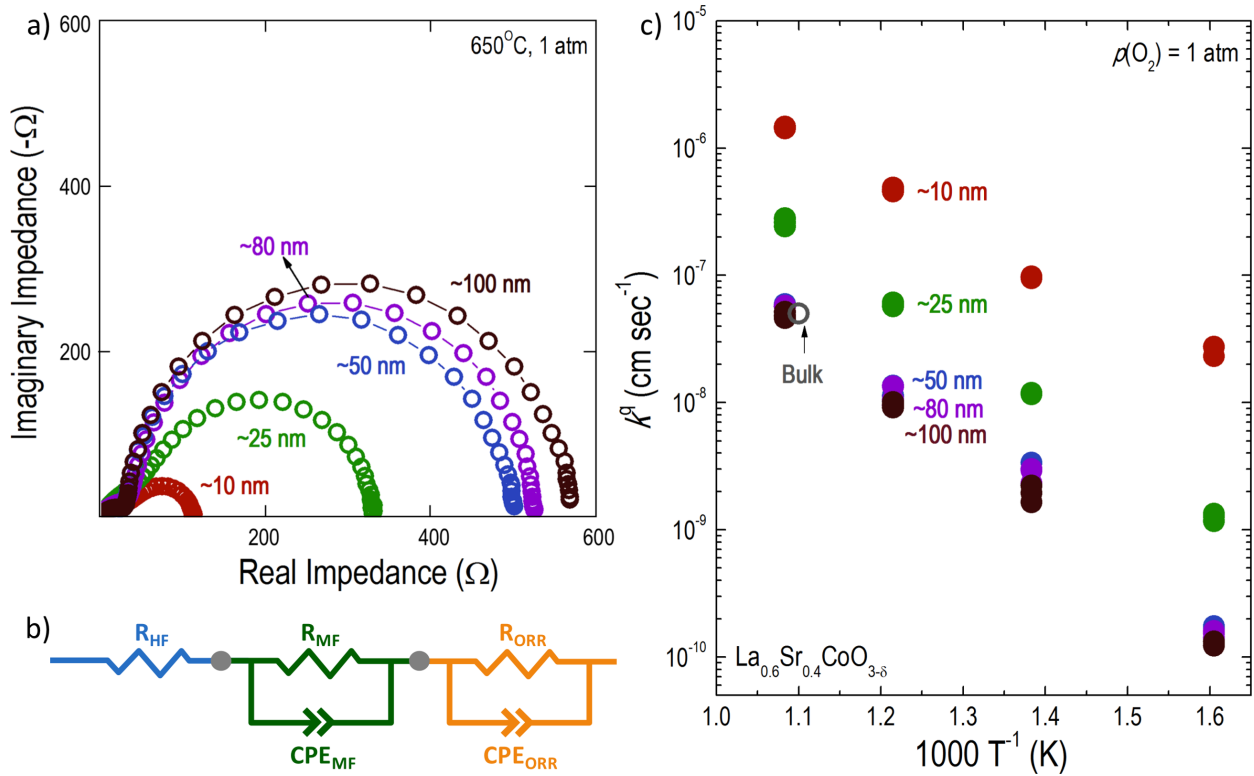

Figure 2. (a) Nyquist plot of LSCO thin films with various thicknesses at 650 °C in 1 atm. (b) Equivalent circuit ($R_{HF}$ = YSZ electrolyte resistance, $R_{MF}$ = electrode/electrolyte interface resistance, $R_{ORR}$ = ORR resistance, CPE = constant phase element) used to extract the ORR kinetics. (c) Temperature dependence of the oxygen surface exchange coefficient ($k^q$) of the LSCO

films calculated from EIS spectra collected at 1 atm. Each sample was measured at least three times at each temperature to ensure data reproducibility. The $k^*$ value of bulk LSCO at 650 °C obtained from literature[32] is also plotted for comparison.

To investigate the ORR activity of the LSCO films, high-temperature electrochemical impedance spectroscopy (EIS) measurements were conducted. Details of EIS data analysis can be found in the Supporting Information. Representative EIS data collected from LSCO thin films with thicknesses of 10, 25, 50, 80, and 100 nm measured at 650 °C with a $p(O_2)$ of 1 atm are shown in Figure 2a. The real part of the impedance with a predominant semicircle decreased significantly with decreasing thickness (or increasing the in-plane tensile strain) of LSCO films. As the thickness of LSCO films is considerably smaller than the critical thickness for bulk transport limitation (~900 μm for bulk LSCO[32] at 600 °C in 1 atm), we conclude that the ORR kinetics of these films is governed primarily by the oxygen surface exchange kinetics, which is further supported by the nearly perfect semicircle impedance curves[33] (Figure 2a) and the $p(O_2)$ dependent impedance responses[34] (Figure S5a).

The EIS data were analyzed using a simplified equivalent circuit shown in Figure 2b. Figure 2c shows the oxygen surface exchange coefficients ($k^q$) of LSCO thin films with different thicknesses as a function of temperature at a $p(O_2)$ of 1 atm. The $k^q$ values of these LSCO thin films were found to decrease with increasing LSCO thickness up to 50 nm, whereas there was no significant change in the $k^q$ value of LSCO films thicker than 50 nm, in which the strain was relaxed. Considering that $k^q$ can be approximated as $k^*$,[35] the $k^q$ value of a highly tensile-strained LSCO (~0.8 % in tensile strain) was dramatically enhanced by approximately two orders of magnitude compared to bulk LSCO.[32] On the other hand, the strain-relaxed LSCO films were found to have comparable $k^q$ values with the $k^*$ value of bulk LSCO.[32] This result suggests that the enhanced ORR kinetics can be attributed to the tensile strain.

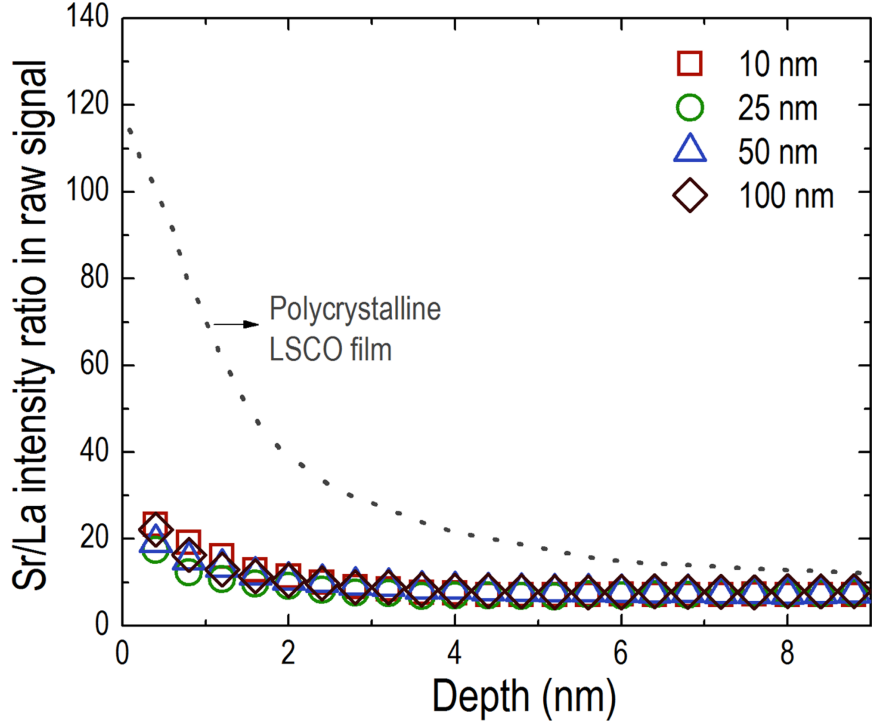

Figure 3. Secondary cation $Sr^+/La^+$ intensity ratio of LSCO films with thicknesses of 10, 25, 50, and 100 nm from ToF-SIMS measurements after annealing at 650 °C for 6 hours. A secondary ion ratio of $Sr^+/La^+$ from a polycrystalline LSCO thin film (~100 nm) extracted from the previous data[36] is also plotted for comparison.

It is known that $La_{1-x}Sr_xCoO_{3-\delta}$ suffers from a reduction in the ORR kinetics associated with the formation of a Sr-enriched layer near the film surface after a heat treatment.[26, 36-38] While there has been no report on the effect of strain on the Sr segregation, the change in the oxygen surface exchange kinetics of LSCO films may be attributed to a change in the degree of Sr segregation near the surface owing to strain and/or thickness modulation. To address this possibility, we performed depth profiling of our LSCO films (Figure S7) after annealing at 650 °C for 6 hours using time-of-flight secondary ion mass spectrometry (ToF-SIMS). Note that the samples were annealed under the same condition as EIS testing to unambiguously check the annealing-induced change in the Sr concentration. Figure 3 shows the secondary ion ratios of $Sr^+/La^+$ of LSCO films (10, 25, 50, and 100 nm in thickness). While a slight increase in the $Sr^+/La^+$ ratio was observed near the top surface (1 - 4 nm) of all LSCO films, the degree of surface Sr enrichment in our LSCO films was not as large as previously reported from thin films with low crystallinity or polycrystalline phase.[36] The reason for the reduced amount of Sr segregation in our study could be the single crystalline nature of our epitaxial films. The grain boundaries commonly found in the polycrystalline films used in other studies may enable more facile Sr transport and therefore greater enrichment of Sr at the surface. The consistent Sr segregation observed in this study across films demonstrates that we can exclude variation in surface Sr enrichment as a cause for the different surface exchange kinetics we are observing in our LSCO films. Moreover, no discernible change in the (Sr+La)/Co ratios of LSCO thin films was observed regardless of the film thickness (Figure S9). Therefore, we can also exclude the effect of the overall changes of the cation ratio in our LSCO films on the observed oxygen surface exchange kinetics.

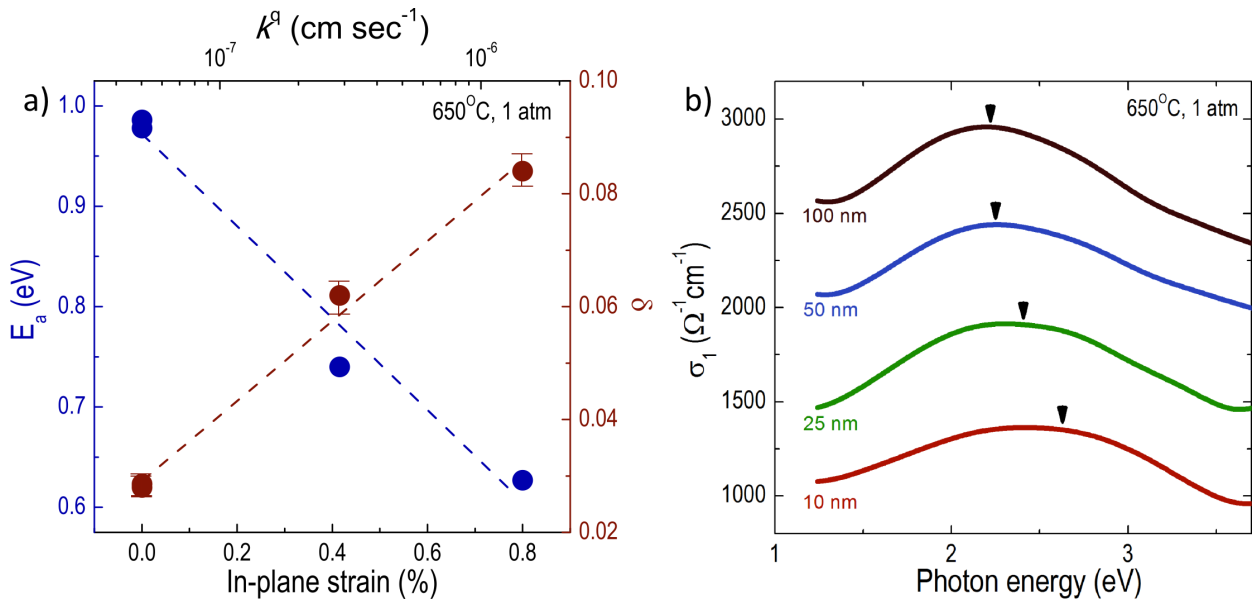

Figure 4. (a) Activation energy (blue) and δ (red) of LSCO films obtained from EIS spectra and high temperature XRD, respectively, measured at 650 °C in 1 atm as a function of in-plane strain and oxygen surface exchange coefficients. For the calculation of δ, a change in the lattice volume

due to the elastic deformation originating from tensile strain in LSCO films was also taken into account. (b) Optical conductivity ($\sigma_1$) of LSCO films with thickness of 10, 25, 50, and 100 nm collected from spectroscopic ellipsometry at 650 $^O$C in 1 atm.

As shown in Figure 4a, manipulating tensile strain through controlling the film thickness causes a considerable reduction in the activation energy barrier for oxygen surface exchange. While the unstrained LSCO films showed a comparable activation energy (~0.98 eV) to the previously reported value from LSCO films,[39-40] the activation energy decreased monotonically with increasing tensile strain in LSCO films (<50 nm in thickness). This observation clearly indicates that a strain-induced reduction in the activation barrier for oxygen surface exchange facilitates the enhanced ORR kinetics of LSCO films.

In addition, in-plane tensile strain is known to create oxygen vacancies in perovskite oxides.[7, 31, 41-42] As shown in Figure 4a, we also observed an increase of oxygen nonstoichiometry ($\delta$) in LSCO films. Note that $\delta$ values in Figure 4a were calculated from unit cell volumes from XRD scans at elevated temperatures according to the established formula[29] (see the Supporting Information). The result of tensile strain increasing the oxygen vacancy concentration is in good agreement with previous computational and experimental reports.[7, 31, 41-42] This result is further supported by the larger unit cell volume at the larger strain (Figure 1c) as an increase in $\delta$ can lead to an increase in the unit cell volume in cobaltite based perovskites.[12, 31] In general, oxygen vacancies provide more active sites for oxygen surface exchange, and thus may also contribute to the enhanced ORR activities in thinner (strained) films (Figure 4a). The relative contribution of vacancies to the reduction in measured activation barrier with strain is discussed later.

Since a change in $\delta$ can be sensitively probed by optical spectroscopy,[43] we measured optical conductivity ($\sigma_1$) by spectroscopic ellipsometry at the EIS measurement condition, i.e., 650

°C and a $p(O_2)$ of 1 atm. Figure 4b shows $\sigma_1$ from LSCO films with various thicknesses. A single peak was observed in the photon energy range between 2 and 3 eV for LSCO films. The peak position was shifted towards lower photon energy with increasing film thickness up to 50 nm, whereas LSCO films with thicknesses above 50 nm showed no difference in the peak position. This result is consistent with a previously reported Sr-doping dependence of the optical conductivity in doped-LaCoO$_3$ at room temperature.[44] Thus, tensile strain results in an increase in oxygen vacancies (i.e., larger $\delta$) as confirmed by the spectroscopic peak shift. This trend is further supported by changes in the volume specific capacitances (VSCs). The VSCs of LSCO thin films were extracted from EIS data (for details see the Supporting Information), which correspond to changes in the oxygen nonstoichiometry ($\delta$) induced by changes in the electrical potential. The VSCs of LSCO thin films were found to decrease with increasing film thickness up to 50 nm, whereas no changes were found above 50 nm, as shown in Figure S6.

As discussed above, the measured reduction in activation barrier for $k^q$ with strain has contributions stemming from the change in $\delta$ as well as changes to the energy barrier for the purely chemical component of the surface exchange processes (e.g. rates of O$_2$ splitting and O surface diffusion). We have used the data of $\delta$ and measured strain-dependent enhancement of $k^q$ to analyze the relative contributions of changes in the oxygen vacancy concentration ($c_{vac}$, where $c_{vac} = \delta/3$) and changes in the vacancy-independent surface exchange factor, $k'$, to the overall enhancement in measured $k^q$ at different strain states using the relation $k^q = c_{vac}^n k'$, with $n = 0$, 1 and 2 (see the Supporting Information for this analysis).[45] If one assumes that the mechanism of surface exchange in LSCO is dissociative adsorption, as found in the work of Adler et al.,[45-47] then $n = 2$. The obtained $p(O_2)$ dependence of $k^q$ for LSCO films is in good agreement with our assumption, where dissociative adsorption was proposed as the rate-limiting step for the oxygen surface exchange in LSCO thin films (see Figure S5b in the Supporting Information). We note that here the term "dissociative adsorption", as discussed in previous studies,[45-46] is the chemical process of splitting the $O_2$ molecule into 2O with the subsequent incorporation into the perovskite lattice via two O vacancies. The relative contributions of $c_{vac}$ and $k'$ to the measured enhancement of $k^q$ are tabulated in Table S1 of the Supporting Information, and plotted in Figure 5. Considering the case of $n = 2$ for dissociative adsorption, then for the strain states of 0.42% and 0.80%, the increase in $c_{vac}$ with strain is the predominant factor in the measured enhancement of $k^q$ (Figure 5). Even in the case of $9.8 \times 10^{-3}$% strain, which is very close to the thick film, unstrained limit, the contribution of vacancies to the $k^q$ enhancement is nearly 50%. Overall, considering a dissociative adsorption surface exchange mechanism, strain influences the value of $c_{vac}$ more than $k'$ for the full range of strain investigated in this work, especially at higher strain states.

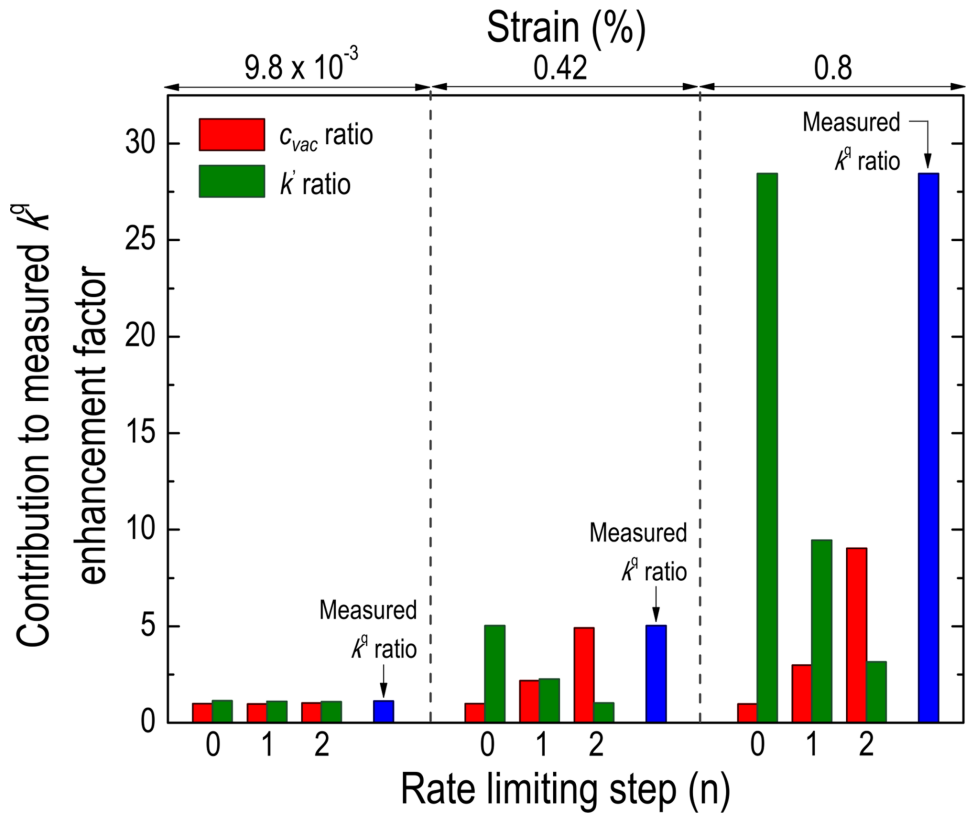

Figure 5. Plot of contributions of $c_{vac}$ and $k'$ to the change in measured $k^q$ as a function of film strain state. Each contribution is plotted relative to the measured value at zero strain (i.e., the values for the film with thickness of 100 nm), for a total measured $k^q$ enhancement of about 28.4× at 0.8% strain (i.e., the values for the film with thickness of 10 nm). The red, green, and blue bars represent the contribution to measured $k^q$ enhancement from vacancies, the contribution from $k'$, and the total change in the measured $k^q$, respectively. All values were obtained by assuming $k^q$ scaled with vacancy concentration as $k^q = c_{vac}^n k'$, where $n = 0$ (reaction limited by, e.g, O$_2$ arrival and is independent of $c_{vac}$), $n = 1$ (reaction is limited by O incorporation and depends linearly on $c_{vac}$), or $n = 2$ (reaction is limited by dissociative adsorption and depends quadratically on $c_{vac}$). Additional details of how these values were calculated and a table of the values are provided in the Supporting Information.

Recent studies for a wide range of bulk perovskites demonstrated that the position of the oxygen 2$p$ band center with respect to the Fermi level is linearly correlated with the activation barrier of oxygen surface exchange, acting as a descriptor for the oxygen surface exchange kinetics.[25, 48] Furthermore, a significant reduction in the migration barrier for oxygen transport was demonstrated under tensile strain in ABO$_3$ perovskites.[6] However, to our knowledge, there has been no study on the correlation between the strain state and the oxygen 2$p$ band center reported in the literature. To explore the relationship among strain, electronic structure, and oxygen surface exchange, we computationally modeled strained LSCO using density functional theory (DFT) calculations for a range of strain states from -3 to +3% (see the Supporting Information for calculation details). We used the change in the oxygen 2$p$ band center relative to the Fermi level as a descriptor.

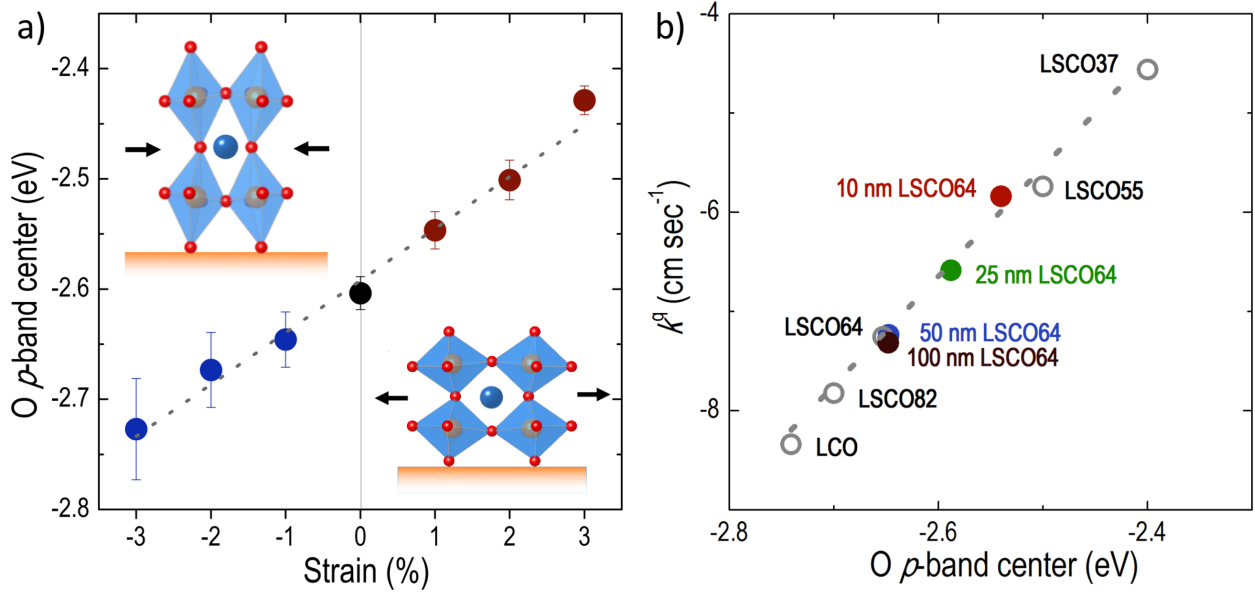

Figure 6. (a) DFT calculations for the oxygen 2$p$ band center with respect to the Fermi level as a function of strain in LSCO films. The calculated oxygen 2$p$ band centers are averaged over LSCO films with five different Sr orderings. (b) Linear relationship between the experimental $k^q$ values and the oxygen 2$p$ band center of LSCO. Bulk $k^q$ values of LSCO (Sr = 0, 0.2, 0.4, 0.5, and 0.7) extracted from previous results[32, 49-51] are also plotted for comparison.

As shown in Figure 6a, the calculated oxygen 2$p$ band center relative to the Fermi level was found to be linearly correlated with strain state, where tensile strain resulted in the upshift of the oxygen 2$p$ band center. As the oxygen 2$p$ band center also has a linear relationship with the vacancy formation energy in perovskite oxides,[25] the tensile strain is expected to lower the energy of vacancy formation due to an increase in the oxygen 2$p$ band center in the LSCO films, which is strongly supported by the fact that larger tensile strains produce more oxygen vacancies as discussed above. This result is also consistent with previous experiments[8, 41] and calculations.[41, 52] In addition, from our above analysis of the contribution of vacancies to the measured enhancement of $k^q$, the vacancy changes are the predominant factor enhancing $k^q$, especially at higher strain states, with the enhancement of $k'$ playing a more minor role.

Figure 6b shows the relationship between the calculated oxygen 2$p$ band center relative to the Fermi level and the oxygen surface exchange kinetics of strained LSCO thin films and bulk LSCO extracted from the references.[32, 49-51] The linear trend, where an oxygen 2$p$ band center value closer to the Fermi level results in an increase of the oxygen surface exchange kinetics demonstrates that the oxygen 2$p$ band center modulated by strain is as an effective descriptor to predict the surface activity, consistent with previously reported data.[25, 48] From Figure 6b, a small upshift of the oxygen 2$p$ band center (on the order of 100 meV) in the tensile-strained LSCO films was found to correlate with the enhancement of $k^q$ values by up to two orders of magnitude, due to the reduced activation energy for $k^q$. This interplay in the change of the electronic structure (upshift of the oxygen 2$p$ band center) and oxygen surface exchange kinetics of our LSCO films strongly supports our conclusion that strain controlled by film thickness enhances the ORR kinetics due to the more facile creation of oxygen vacancies, thus increasing the oxygen vacancy concentration and subsequently increasing the oxygen surface exchange rate.

CONCLUSIONS

In summary, we successfully demonstrated a significantly enhanced ORR activity in epitaxial LSCO thin films on YSZ by thickness-controlled strain. We showed that a small change in strain has a negligible effect on the surface Sr concentration, indicating no influence of the Sr surface segregation on the ORR kinetics. We revealed that the enhanced ORR activity is strongly associated with a change in the electronic structure, yielding an increase in the oxygen vacancy concentration, thus increasing the oxygen surface exchange kinetics by tensile strain. These findings demonstrate the key role of epitaxial strain in the oxygen surface exchange process, which can provide a new avenue to designing high performance energy materials for clean energy conversion and storage devices.

ASSOCIATED CONTENT

Supporting Information

The Supporting Information is available free of charge on the ACS Publications website at DOI:

Additional figures, details of film growth, in-situ XRD analysis, EIS data analysis, lattice parameter calculations, strain analysis, unit cell volume calculations, nonstoichiometry calculations, DFT calculations, and activation barrier analysis.

AUTHOR INFORMATION

Corresponding Author

*E-mail: hnlee@ornl.gov

Notes

The authors declare no competing financial interest.

ACKNOWLEDGMENT

This work was supported by the U.S. Department of Energy (DOE), Office of Science, Basic Energy Sciences, Materials Science and Engineering Division (synthesis and structural characterization) and by the Laboratory Directed Research and Development Program of Oak Ridge National Laboratory (ORNL), managed by UT-Battelle, LLC, for the U. S. DOE (electrochemical characterization). The high temperature XRD and SIMS measurements were conducted as a user project at the Center for Nanophase Materials Sciences, which is sponsored at ORNL by the Scientific User Facilities Division, U.S. DOE. Support for Ryan Jacobs and Dane Morgan for DFT calculations was provided by the National Science Foundation (NSF) Software Infrastructure for Sustained Innovation (SI2) award No. 1148011. Computations in this work benefitted from the use of the Extreme Science and Engineering Discovery Environment (XSEDE), which is supported by NSF grant number OCI-1053575. We gratefully acknowledge helpful conversations with Professor Stu Adler.

REFERENCES

(1) Gasteiger, H. A.; Marković, N. M. Just a Dream or Future Reality? *Science* **2009**, *324*, 48-49.

(2) Hashim, S. M.; Mohamed, A. R.; Bhatia, S. Current Status of Ceramic-Based Membranes for Oxygen Separation from Air. *Adv. Colloid Interface Sci.* **2010**, *160*, 88-100.

(3) Suntivich, J.; Gasteiger, H. A.; Yabuuchi, N.; Nakanishi, H.; Goodenough, J. B.; Shao-Horn, Y. Design Principles for Oxygen-Reduction Activity on Perovskite Oxide Catalysts for Fuel Cells and Metal-Air Batteries. *Nat. Chem.* **2011**, *3*, 647-647.

(4) Shao, Z. P.; Haile, S. M. A High-Performance Cathode for the Next Generation of Solid-Oxide Fuel Cells. *Nature* **2004**, *431*, 170-173.

(5) Cai, Z. H.; Kuru, Y.; Han, J. W.; Chen, Y.; Yildiz, B. Surface Electronic Structure Transitions at High Temperature on Perovskite Oxides: The Case of Strained $La_{0.8}Sr_{0.2}CoO_3$ Thin Films. *J. Am. Chem. Soc.* **2011**, *133*, 17696-17704.

(6) Mayeshiba, T.; Morgan, D. Strain Effects on Oxygen Migration in Perovskites. *Phys. Chem. Chem. Phys.* **2015**, *17*, 2715-2721.

(7) Aschauer, U.; Pfenninger, R.; Selbach, S. M.; Grande, T.; Spaldin, N. A. Strain-Controlled Oxygen Vacancy Formation and Ordering in $CaMnO_3$. *Phys. Rev. B* **2013**, *88*, 054111.

(8) Kubicek, M.; Cai, Z. H.; Ma, W.; Yildiz, B.; Hutter, H.; Fleig, J. Tensile Lattice Strain Accelerates Oxygen Surface Exchange and Diffusion in $La_{1-x}Sr_xCoO_{3-\delta}$ Thin Films. *ACS Nano* **2013**, *7*, 3276-3286.

(9) la O, G. J.; Ahn, S. J.; Crumlin, E.; Orikasa, Y.; Biegalski, M. D.; Christen, H. M.; Shao-Horn, Y. Catalytic Activity Enhancement for Oxygen Reduction on Epitaxial Perovskite Thin Films for Solid-Oxide Fuel Cells. *Angew. Chem. Int. Ed.* **2010**, *49*, 5344-5347.

(10) Lee, D.; Grimaud, A.; Crumlin, E. J.; Mezghani, K.; Habib, M. A.; Feng, Z. X.; Hong, W. T.; Biegalski, M. D.; Christen, H. M.; Shao-Horn, Y. Strain Influence on the Oxygen Electrocatalysis of the (100)-Oriented Epitaxial $La_2NiO_{4+\delta}$ Thin Films at Elevated Temperatures. *J. Phys. Chem. C* **2013**, *117*, 18789-18795.

(11) Crumlin, E. J.; Ahn, S. J.; Lee, D.; Mutoro, E.; Biegalski, M. D.; Christen, H. M.; Shao-Horn, Y. Oxygen Electrocatalysis on Epitaxial $La_{0.6}Sr_{0.4}CoO_{3-\delta}$ Perovskite Thin Films for Solid Oxide Fuel Cells. *J. Electrochem. Soc.* **2012**, *159*, F219-F225.

(12) Hong, W. T.; Gadre, M.; Lee, Y.-L.; Biegalski, M. D.; Christen, H. M.; Morgan, D.; Shao-Horn, Y. Tuning the Spin State in $LaCoO_3$ Thin Films for Enhanced High-Temperature Oxygen Electrocatalysis. *J. Phys. Chem. Lett.* **2013**, *4*, 2493-2499.

(13) Mitterdorfer, A.; Gauckler, L. J. $La_2Zr_2O_7$ Formation and Oxygen Reduction Kinetics of the $La_{0.85}Sr_{0.15}Mn_yO_3$, $O_2(g)$│YSZ System. *Solid State Ion.* **1998**, *111*, 185-218.

(14) Ma, W.; Kim, J. J.; Tsvetkov, N.; Daio, T.; Kuru, Y.; Cai, Z. H.; Chen, Y.; Sasaki, K.; Tuller, H. L.; Yildiz, B. Vertically Aligned Nanocomposite $La_{0.8}Sr_{0.2}CoO_3/(La_{0.5}Sr_{0.5})_2CoO_4$ Cathodes - Electronic Structure, Surface Chemistry and Oxygen Reduction Kinetics. *J. Mater. Chem. A* **2015**, *3*, 207-219.

(15) Kresse, G.; Furthmuller, J. Efficient Iterative Schemes for ab Initio Total-Energy Calculations Using a Plane-Wave Basis Set. *Phys. Rev. B* **1996**, *54*, 11169-11186.

(16) Hubbard, J. Electron Correlations in Narrow Energy Bands. *Proc. R. Soc. London, Ser. A* **1963**, *276*, 238-257.

(17) Kresse, G.; Joubert, D. From Ultrasoft Pseudopotentials to the Projector Augmented-Wave Method. *Phys. Rev. B* **1999**, *59*, 1758-1775.

(18) Perdew, J. P.; Wang, Y. Accurate and Simple Analytic Representation of the Electron-Gas Correlation Energy. *Phys. Rev. B* **1992**, *45*, 13244-13249.

(19) Wang, L.; Maxisch, T.; Ceder, G. Oxidation Energies of Transition Metal Oxides within the GGA+U Framework. *Phys. Rev. B* **2006**, *73*, 195107.

(20) Pandey, S. K.; Kumar, A.; Patil, S.; Medicherla, V. R. R.; Singh, R. S.; Maiti, K.; Prabhakaran, D.; Boothroyd, A. T.; Pimpale, A. V. Investigation of the Spin State of Co in $LaCoO_3$ at Room Temperature: ab Initio Calculations and High-Resolution Photoemission Spectroscopy of Single Crystals. *Phys. Rev. B* **2008**, *77*, 045123.

(21) Gadre, M. Ph.D. Thesis. University of Wisconsin- Madison, 2014.

(22) Monkhorst, H. J.; Pack, J. D. Special Points for Brillouin-Zone Integrations. *Phys. Rev. B* **1976**, *13*, 5188-5192.

(23) Jacobs, R.; Booske, J.; Morgan, D. Understanding and Controlling the Work Function of Perovskite Oxides Using Density Functional Theory. *Adv. Funct. Mater.* **2016**, *26*, 5471-5482.

(24) Lee, Y.-L.; Kleis, J.; Rossmeisl, J.; Morgan, D. ab Initio Energetics of La*B*$O_3$ (001) (*B* = Mn, Fe, Co, and Ni) for Solid Oxide Fuel Cell Cathodes. *Phys. Rev. B* **2009**, *80*, 224101.

(25) Lee, Y.-L.; Kleis, J.; Rossmeisl, J.; Shao-Horn, Y.; Morgan, D. Prediction of Solid Oxide Fuel Cell Cathode Activity with First-Principles Descriptors. *Energy Environ. Sci.* **2011**, *4*, 3966-3970.

(26) Lee, D.; Lee, Y.-L.; Hong, W. T.; Biegalski, M. D.; Morgan, D.; Shao-Horn, Y. Oxygen Surface Exchange Kinetics and Stability of $(La,Sr)_2CoO_{4\pm\delta}/La_{1-x}Sr_xMO_{3-\delta}$ (*M* = Co and Fe) Hetero-Interfaces at Intermediate Temperatures. *J. Mater. Chem. A* **2015**, *3*, 2144-2157.

(27) Sonntag, R.; Neov, S.; Kozhukharov, V.; Neov, D.; ten Elshof, J. E. Crystal and Magnetic Structure of Substituted Lanthanum Cobalitites. *Physica B* **1997**, *241*, 393-396.

(28) Yashima, M.; Sasaki, S.; Kakihana, M.; Yamaguchi, Y.; Arashi, H.; Yoshimura, M. Oxygen-Induced Structural Change of the Tetragonal Phase around the Tetragonal-Cubic Phase-Boundary in $ZrO_2$-$YO_{1.5}$ Solid Solutions. *Acta Cryst.* **1994**, *50*, 663-672.

(29) Chen, X. Y.; Yu, J. S.; Adler, S. B. Thermal and Chemical Expansion of Sr-Doped Lanthanum Cobalt Oxide ($La_{1-x}Sr_xCoO_{3-\delta}$). *Chem. Mater.* **2005**, *17*, 4537-4546.

(30) Ishihara, T.; Kudo, T.; Matsuda, H.; Takita, Y. Doped $PrMnO_3$ Perovskite Oxide as a New Cathode of Solid Oxide Fuel Cells for Low Temperature Operation. *J. Electrochem. Soc.* **1995**, *142*, 1519-1524.

(31) Petrie, J. R.; Mitra, C.; Jeen, H.; Choi, W. S.; Meyer, T. L.; Reboredo, F. A.; Freeland, J. W.; Eres, G.; Lee, H. N. Strain Control of Oxygen Vacancies in Epitaxial Strontium Cobaltite Films. *Adv. Funct. Mater.* **2016**, *26*, 1564-1570.

(32) Berenov, A. V.; Atkinson, A.; Kilner, J. A.; Bucher, E.; Sitte, W. Oxygen Tracer Diffusion and Surface Exchange Kinetics in $La_{0.6}Sr_{0.4}CoO_{3-\delta}$. *Solid State Ion.* **2010**, *181*, 819-826.

(33) Adler, S. B. Factors Governing Oxygen Reduction in Solid Oxide Fuel Cell Cathodes. *Chem. Rev.* **2004**, *104*, 4791-4843.

(34) Adler, S. B.; Lane, J. A.; Steele, B. C. H. Electrode Kinetics of Porous Mixed-Conducting Oxygen Electrodes. *J. Electrochem. Soc.* **1996**, *143*, 3554-3564.

(35) Maier, J. On the Correlation of Macroscopic and Microscopic Rate Constants in Solid State Chemistry. *Solid State Ion.* **1998**, *112*, 197-228.

(36) Kubicek, M.; Limbeck, A.; Fromling, T.; Hutter, H.; Fleig, J. Relationship between Cation Segregation and the Electrochemical Oxygen Reduction Kinetics of $La_{0.6}Sr_{0.4}CoO_{3-\delta}$ Thin Film Electrodes. *J. Electrochem. Soc.* **2011**, *158*, B727-B734.

(37) Crumlin, E. J.; Mutoro, E.; Liu, Z.; Grass, M. E.; Biegalski, M. D.; Lee, Y.-L.; Morgan, D.; Christen, H. M.; Bluhm, H.; Shao-Horn, Y. Surface Strontium Enrichment on Highly Active Perovskites for Oxygen Electrocatalysis in Solid Oxide Fuel Cells. *Energy Environ. Sci.* **2012**, *5*, 6081-6088.

(38) Lee, D.; Lee, Y.-L.; Grimaud, A.; Hong, W. T.; Biegalski, M. D.; Morgan, D.; Shao-Horn, Y. Enhanced Oxygen Surface Exchange Kinetics and Stability on Epitaxial $La_{0.8}Sr_{0.2}CoO_{3-\delta}$ Thin Films by $La_{0.8}Sr_{0.2}MnO_{3-\delta}$ Decoration. *J. Phys. Chem. C* **2014**, *118*, 14326-14334.

(39) Januschewsky, J.; Ahrens, M.; Opitz, A.; Kubel, F.; Fleig, J. Optimized $La_{0.6}Sr_{0.4}CoO_{3-\delta}$ Thin-Film Electrodes with Extremely Fast Oxygen-Reduction Kinetics. *Adv. Funct. Mater.* **2009**, *19*, 3151-3156.

(40) Yang, Y. L.; Chen, C. L.; Chen, S. Y.; Chu, C. W.; Jacobson, A. J. Impedance Studies of Oxygen Exchange on Dense Thin Film Electrodes of $La_{0.5}Sr_{0.5}CoO_{3-\delta}$. *J. Electrochem. Soc.* **2000**, *147*, 4001-4007.

(41) Donner, W.; Chen, C. L.; Liu, M.; Jacobson, A. J.; Lee, Y.-L.; Gadre, M.; Morgan, D. Epitaxial Strain-Induced Chemical Ordering in $La_{0.5}Sr_{0.5}CoO_{3-\delta}$ Films on $SrTiO_3$. *Chem. Mater.* **2011**, *23*, 984-988.

(42) Petrie, J. R.; Jeen, H.; Barron, S. C.; Meyer, T. L.; Lee, H. N. Enhancing Perovskite Electrocatalysis through Strain Tuning of the Oxygen Deficiency. *J. Am. Chem. Soc.* **2016**, *138*, 7252-7255.

(43) Choi, W. S.; Kwon, J.-H.; Jeen, H.; Hamann-Borrero, J. E.; Radi, A.; Macke, S.; Sutarto, R.; He, F.; Sawatzky, G. A.; Hinkov, V.; Kim, M. et al. Strain-Induced Spin States in Atomically Ordered Cobaltites. *Nano Lett.* **2012**, *12*, 4966-4970.

(44) Yamaguchi, S.; Okimoto, Y.; Ishibashi, K.; Tokura, Y. Magneto-Optical Kerr Effects in Perovskite-Type Transition-Metal Oxides: $La_{1-x}Sr_xMnO_3$ and $La_{1-x}Sr_xCoO_3$. *Phys. Rev. B* **1998**, *58*, 6862-6870.

(45) Lu, Y.; Kreller, C.; Adler, S. B. Measurement and Modeling of the Impedance Characteristics of Porous $La_{1-x}Sr_xCoO_{3-\delta}$ Electrodes. *J. Electrochem. Soc.* **2009**, *156*, B513-B525.

(46) Adler, S. B.; Chen, X. Y.; Wilson, J. R. Mechanisms and Rate Laws for Oxygen Exchange on Mixed-Conducting Oxide Surfaces. *J. Catal.* **2007**, *245*, 91-109.

(47) Wilson, J. R.; Sase, M.; Kawada, T.; Adler, S. B. Measurement of Oxygen Exchange Kinetics on Thin-Film $La_{0.6}Sr_{0.4}CoO_{3-\delta}$ using Nonlinear Electrochemical Impedance Spectroscopy. *Electrochem. Solid-State Lett.* **2007**, *10*, B81-B86.

(48) Lee, Y.-L.; Lee, D.; Wang, X. R.; Lee, H. N.; Morgan, D.; Shao-Horn, Y. Kinetics of Oxygen Surface Exchange on Epitaxial Ruddlesden–Popper Phases and Correlations to First-Principles Descriptors. *J. Phys. Chem. Lett.* **2016**, *7*, 244-249.

(49) De Souza, R. A.; Kilner, J. A. Oxygen Transport in $La_{1-x}Sr_xMn_{1-y}Co_yO_{3\pm\delta}$ Perovskites Part II. Oxygen Surface Exchange. *Solid State Ion.* **1999**, *126*, 153-161.

(50) Ishigaki, T.; Yamauchi, S.; Mizusaki, J.; Fueki, K.; Tamura, H. Tracer Diffusion-Coefficient of Oxide Ions in $LaCoO_3$ Single-Crystal. *J. Solid State Chem.* **1984**, *54*, 100-107.

(51) van Doorn, R. H. E.; Fullarton, I. C.; de Souza, R. A.; Kilner, J. A.; Bouwmeester, H. J. M.; Burggraaf, A. J. Surface Oxygen Exchange of $La_{0.3}Sr_{0.7}CoO_{3-\delta}$. *Solid State Ion.* **1997**, *96*, 1-7.

(52) Han, J. W.; Yildiz, B. Enhanced One Dimensional Mobility of Oxygen on Strained $LaCoO_3$ (001) Surface. *J. Mater. Chem.* **2011**, *21*, 18983-18990.

Table of Contents (TOC)

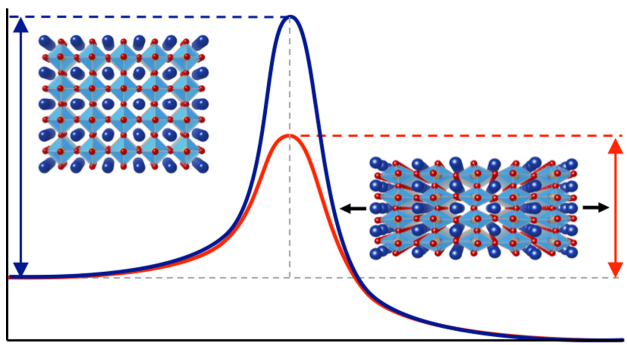